\documentclass[twocolumn]{aastex63}

\usepackage{array, booktabs}
\usepackage{tablefootnote}
\usepackage{footnotehyper}
\usepackage{ulem}
\usepackage{gensymb}
\usepackage{graphicx}

\newcommand{\foo}{\makebox[0pt]{\textbullet}\hskip-0.5pt\vrule width 1pt\hspace{\labelsep}}

\newcommand{\pcm}{\,cm$^{-2}$}	% per cm-squared
\newcommand{\swift}{\textit{Swift}}
\newcommand{\fermi}{\textit{Fermi}}

\shorttitle{GUANO: BAT event data on demand}
\shortauthors{Tohuvavohu et al.}
\graphicspath{{./}{figures/}}
\begin{document}

\title{Gamma-ray Urgent Archiver for Novel Opportunities (GUANO): Swift/BAT event data dumps on demand to enable sensitive sub-threshold GRB searches}

\correspondingauthor{Aaron Tohuvavohu}
\email{tohuvavohu@astro.utoronto.ca}

\author[0000-0002-2810-87647]{Aaron Tohuvavohu}
\affiliation{Department of Astronomy and Astrophysics, University of Toronto, Toronto, ON, CA}

\author[0000-0002-6745-4790]{Jamie A. Kennea}
\affiliation{Department of Astronomy \& Astrophysics, Penn State University, PA, USA}

\author[0000-0001-5229-1995]{James DeLaunay}
\affiliation{Department of Astronomy \& Astrophysics, Penn State University, PA, USA}

\author{David M. Palmer}
\affiliation{Los Alamos National Laboratory, Los Alamos, New Mexico, USA}

\author[0000-0003-1673-970X]{S. Bradley Cenko}
\affiliation{NASA Goddard Space Flight Center, Greenbelt, MD, USA}

\author{Scott Barthelmy}
\affiliation{NASA Goddard Space Flight Center, Greenbelt, MD, USA}



\begin{abstract}
We introduce a new capability of the Neil Gehrels \swift\ Observatory, to provide event-level data from the Burst Alert Telescope (BAT) on demand in response to transients detected by other instruments. These event-level data are not continuously available due to the large telemetry load, and limited downlink bandwidth, and are critical to recovering weak and/or sub-threshold GRBs that are not triggered onboard, such as the likely counterparts to GW-detected off-axis binary neutron star mergers. We show that the availability of the event data can effectively increase the rate of detections, and arcminute localizations, of GRB 170817-like bursts by $>400\%$. We describe an autonomous spacecraft commanding pipeline purpose built to enable this science; to our knowledge the first fully autonomous extremely-low-latency commanding of a space telescope for scientific purposes. This pipeline has been successfully run in its complete form since January 2020, and has resulted in the recovery of BAT event data for $>700$ externally triggered events to date (GWs, neutrinos, GRBs triggered by other facilities, FRBs, and VHE detections), now running with a success rate of $\sim90\%$. We exemplify the utility of this new capability by using the resultant data to (1) set the most sensitive (8 sigma) upper limits of $8.1\times10^{-8}$ erg\pcm\ s$^{-1}$ (14-195 keV) on prompt 1s duration short GRB-like emission within +/- 15s around the unmodelled GW burst candidate S200114f, and (2) provide an arcminute localization for short GRB 200325A and other bursts. We also show that using data from GUANO to localize GRBs discovered by other instruments, we can increase the net rate of arcminute localized GRBs by $10-20\%$ per year. Along with the scientific yield of more sensitive searches for sub-threshold GRBs, the new capabilities designed for this project will serve as the foundation for further automation and rapid response Target of Opportunity capabilities for the \swift\ mission, and also have implications for the design of future rapid-response space telescopes.

\end{abstract}

\keywords{Gamma-ray bursts --- gravitational wave sources --- space telescopes --- methods: miscellaneous} 

\section{Introduction} \label{sec:intro}
The Burst Alert Telescope (BAT; \citealt{BAT}) onboard the Neil Gehrels \swift\ Observatory (\swift; \citealt{swift}) is the most sensitive gamma-ray burst (GRB) detector in current operation and the only one  \textit{frequently}\footnote{\textit{INTEGRAL}/IBIS also can localize to arc-minute precision, but its detection rate is $>\sim~15$ lower than BAT.} providing arcminute localizations of hard X-ray and gamma-ray transients. 
BAT has been enormously successful in its main missions 1) detecting and promptly localizing GRBs with arc-minute accuracy \citep{Swiftreview} and 2) performing the most sensitive and highest resolution all sky hard X-ray survey to date \citep{BAT105survey}. 

The overwhelming majority of BAT detected GRBs are found utilizing the onboard triggering algorithms \citep{triggeralgo}, which determine the reality of a triggering event by both comparing the detector event rates, and by constructing a sky image and searching for significantly detected ($>6.5\sigma$) new point sources. 
However, despite the success of the BAT onboard triggering algorithm, BAT can detect GRBs which do not trigger the onboard algorithm. In order to recover these GRBs, we must rely on ground searches that run on downlinked  BAT data to search for these untriggered events. However, the utility of the BAT to search for events which did not trigger onboard (untriggered, or sub-threshold) has been limited due to the unavailability of the event-level data on the ground. This is unfortunate, as despite its limited field of view ($\sim7000$ sq. degrees) in comparison to e.g. \fermi/GBM \citep{FermiGBM}, BAT's superior sensitivity in principle allows access to weak sGRBs that would be otherwise undetectable by other missions, and its arcminute localizing power is critical to rapid follow-up of these events. The unavailability of the event data for ground searches has also necessitated the development of BAT trigger simulators \citep{triggersimulator, triggermodelML} in order to perform population inference of the cosmic GRB population, corrected for the complex selection functions/biases of the BAT onboard triggering algorithms.

In normal operations, BAT records the arrival time (to 100 $\mu$s accuracy), location (in detector coordinates), and energy (in one of 80 bins from 15-300 keV) for each individual count that strikes the detector. This data is referred to as event-by-event (or simply: event) data. Due to the large effective area of the BAT, the data volume produced is quite large, and cannot all be stored onboard or telemetered to the ground. For this reason the BAT has relied on the performance of its onboard real-time detection algorithms, and only preserves event data and telemeters it to the ground around the time of events that trigger these onboard algorithms.

If there is no triggered event for a certain time period, the only BAT data products available for analysis on the ground are summed-array rates light curves (in 64ms, 1s and 1.6s\footnote{The 1.6s binned rates are available as summed counts from quadrants of the full array.} bins in 4 energy channels), 64s single-bin (15-50 keV) images, and 300s 80-bin images for use in the all sky survey. While bright GRBs \textit{can} be identified in the rates data, event mode data are necessary to construct sky images for localization, to remove the hard X-ray and particle background, to identify and remove the effects of glitches or noisy individual detectors, and for the creation of background-subtracted light curves for individual sources. Importantly, the complete event mode data are necessary to reliably distinguish between detector noise and real dim/sub-threshold GRBs. (DeLaunay et al. 2020, in prep).

In order to assess the capabilities of the BAT onboard triggering algorithms vs those achievable if the event data were available on the ground, we choose as an example the one known GW/GRB event to date, GRB 170817 \citep{Abbott_2017}. GRB 170817 was not detected by BAT, as unfortunately the burst location was occulted by the Earth at the time of the GW and GRB (for more details see \citealt{Phil170817}). 

\begin{figure}
    \centering
    \includegraphics[width=0.5\textwidth]{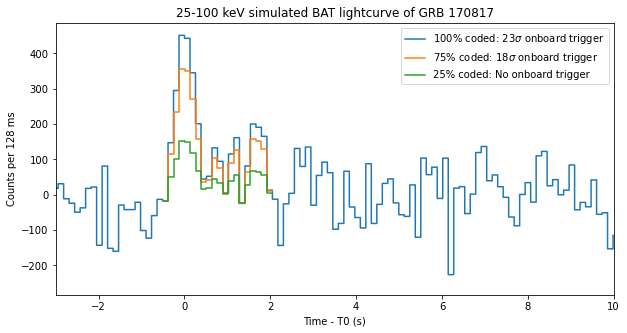}
    \caption{GRB 170817, at its distance of 43 Mpc, simulated in BAT from 25-100 keV, for 3 different angles with respect to the BAT boresight.}
    \label{fig:bat170817}
\end{figure}

We model the light curve shape of GRB 170817, and take the best fit spectral parameters, $E_{\alpha}=-0.62$ and $E_{peak}=185.0$ keV measured from the \fermi/GBM observation \citep{goldstein170817}. Using the BAT Trigger Simulator software \citep{triggersimulator}, and setting the number of active detectors in the BAT array to N=18,000,\footnote{18,000 was chosen as this represents a normal number of active detectors during early 2020. BAT has 32,768 detectors in the array, but over the lifetime of the mission more of these become permanently noisy and are deactivated. See Figure 3 in \citealt{triggersimulator}.} we introduce realistic background on top of the light curve, and then simulate the BAT triggering response as a function of position in the BAT FOV. The simulated BAT light curves, and the triggering result, can be seen in Figure \ref{fig:bat170817}. At its distance of 43 Mpc, GRB170817 would have resulted in an onboard trigger at $>18$ sigma significance if it were $\leq30\deg$ off the boresight of BAT (corresponding to a partial coding of at least $\sim50-60\%$ depending on the BAT's orientation). 

We repeat the exercise, this time simulating a GRB 170817-like burst at various distances to asses the triggering range. The maximum distance at which a GRB 170817-like burst would likely trigger BAT onboard is found to be $\sim$ 67 Mpc\, for the most favorable observing scenario of the burst occurring at 100\% partial coding. This should not be taken as an average range, only a maximum one, as most other locations within the BAT FOV will have reduced ranges compared to this.  To provide context with respect to other GRB missions, GRB 170817 would not trigger \fermi/GBM onboard beyond 50 Mpc \citep{goldstein170817}, and would not be detectable by the GBM targeted search ground analysis beyond $\sim74$ Mpc \citep{GBMsubthresh}. 
 
 However, if the BAT event data were are available on the ground, more sensitive targeted searches can be run around the time of the GW events. 
Using a likelihood-based targeted search (see DeLaunay et al. 2020, in prep and Section \ref{sec:mll} of this work), we find that a GRB 170817-like burst can be recovered out to a distance of $\sim$ 100 Mpc\ if it were at the center of the field of view and out to $\sim$ 71 Mpc\ if it were 45$\deg$ off the boresight of BAT. Weighting the average range achievable over the entire coded field-of-view with the event data available on the ground, vs that from relying on onboard triggers, this range increase corresponds to a volumetric rate increase of $>400\%$ for the detection, and arcminute localization, of GRB 170817-like bursts.
% On-axis:
% 90% recovery, ~93 Mpc
% 50% recovery, ~100 Mpc
% At imx=1.0, imy=0.0 (an optimal 45deg off-axis):
% 90% recovery, ~65 Mpc
% 50% recovery, ~71 Mpc

In addition to dramatically extending the range, and thus volumetric rate, of detections (and importantly arcminute localizations) of GRB 170817-like bursts, the availability of event data on the ground would also increase the rate by a further $\sim15\%$ by correcting for the duty cycle limitations of BAT. \swift\ spends $\sim15\%$ of its time slewing from target to target. During these times, the BAT onboard trigger algorithms are not run. However, GRBs can be found and localized during slews using the event data on the ground (see eg \citealt{BATSS} and Section \ref{sec:grbselsewhere} of this work) with little-to-no decrease in the sensitivity to short bursts. 

The rate enhancements of detection and arcminute localization of GRB 170817-like bursts are only possible with the event data available on the ground, and clearly motivate extraordinary efforts to recover this data. The development in 2012 of the capability to save \textit{all} of the Continuous Time Tagged Event (CTTE) data from the \fermi/GBM instrument, was critical to the development of powerful ground analyses that extend their range for targeted searches, and that now recover an extra $\sim40$ short GRB candidates per year \citep{GBMsubthresh} that do not trigger \fermi/GBM onboard.  Bringing \textit{all} of the event data down is not an option for \swift\ given BAT's higher effective area (and thus count rate) and the fact that, unlike \fermi, \swift\ does not have a high-gain antenna. So we are both telemetry load and bandwidth limited.

Instead, in this paper we describe a newly developed capability for \swift\, to save the event mode data \textit{on demand}, in response to a trigger from an external instrument, and rapidly telemeter it to the ground for use in new powerful targeted sub-threshold GRB searches, especially with application for the search of coincident GW/GRB events.

In sections 3, 4, 5, and 6 we describe relevant technical, design, and implementation details of the GUANO pipeline, and evaluate its performance to date. In Sections 7 and 8 we provide some direct examples of scientific results derived from data recovered by this pipeline. 

\section{BAT Ring Buffer}
\label{sec:ringbuffer}

\begin{figure}
    \includegraphics[width=.5\textwidth]{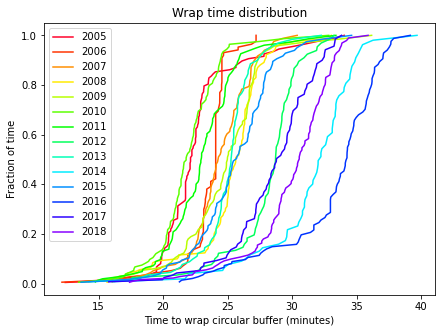}
    \caption{A plot of the distribution of the amount of time it takes to wrap the ring buffer, as a function of year. A general secular trend towards longer wrap times as a function of year is seen (due to more detectors being disabled), but the increasing trend is not strict, due to changing instrument and calibration settings. In 2018, recovering the desired data 99\% of the time would require a command within 20 minutes.}
    \label{fig:wrapduration}
\end{figure}

The event mode data are stored in a ring buffer on the BAT instrument computer, which overwrites itself once it reaches approximately 23 million counts. The effective look-back time (how long before any given piece of event mode data is overwritten) varies on short timescales due to the varying full detector count rate (from varying background levels throughout the orbit, GRBs, bright X-ray sources, and detector noise), and on the timescale of years due to the gradual decrease in detector effective area over time. These varying factors and their impact can be seen in Figure \ref{fig:wrapduration}.

In order to save the event data of interest, it must be moved from the ring buffer to the solid-state recorder (and marked for down-link) before being overwritten, which can only be performed by sending a command to perform this task to \swift. For this reason, extremely low latency commanding of the spacecraft is required in response to an external trigger (GW, neutrino, GRB, etc). 

\section{South Atlantic Anomaly}
During passage of the South Atlantic Anomaly (SAA), $<10\%$ of the time, the BAT instrument does not record information on individual events due to the extremely high count rates. \footnote{
The full array summed, 1-second binned, count rate is recorded through SAA passage, as this is derived from a hardware counter. It is typically $>1$ Million cts/s.} For this reason, any event trigger occurring while \swift\ is passing through the SAA will not have any event data recorded. However, the location of the SAA boundary is not hard coded into the BAT instrument (for \swift's narrow field instruments, and the \fermi\ instruments, the SAA definition is a coordinate defined polygon). Instead BAT determines whether it is in SAA-mode dynamically, determining entry to SAA based on the size of the current onboard data processing backlog, and exit from SAA-mode based on the instantaneous count rate recorded.

\begin{figure}
    \centering
    \includegraphics[width=.5\textwidth]{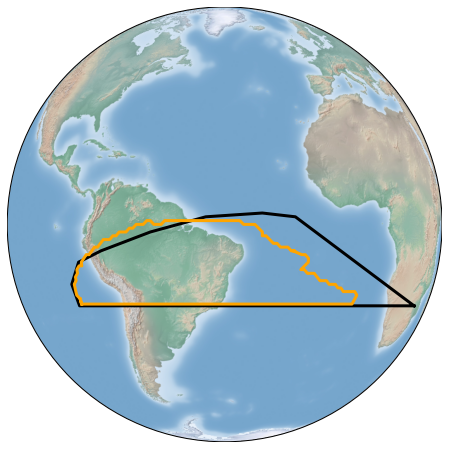}
    \caption{The average spatial extent of the BAT SAA-mode (when BAT does not record event data) in 2019 in orange, compared to the fixed SAA polygon definition used for scheduling observations with the XRT/UVOT in black. The regions are cut off at $21.5\degree$ South, as this is the maximum southern extent of \swift's orbit. BAT is in SAA-mode $\sim8-9\%$ of the time.}
    \label{fig:saa}
\end{figure}

The physical SAA boundary and extent is dynamic on short timescales, responding to space weather and events like Coronal Mass Ejections from the Sun \citep{SAAbyrxte}. In addition to this, using 15 years of \swift\ telemetry, we have determined the average spatial extent of the SAA according to BAT has changed significantly over time, meaning that any definition of the SAA boundary needs to be calculated from recent data.
Therefore for the purposes of the GUANO pipeline, we calculated a region, shown in \ref{fig:saa}, which is used for screening times of triggering events that occurred during SAA passage. This region is defined by the latitude/longitude distribution of \swift\ at times when the BAT was in SAA-mode, during 2019.

\section{GUANO pipeline}
\begin{figure*}
    \centering
    \includegraphics[width=\textwidth]{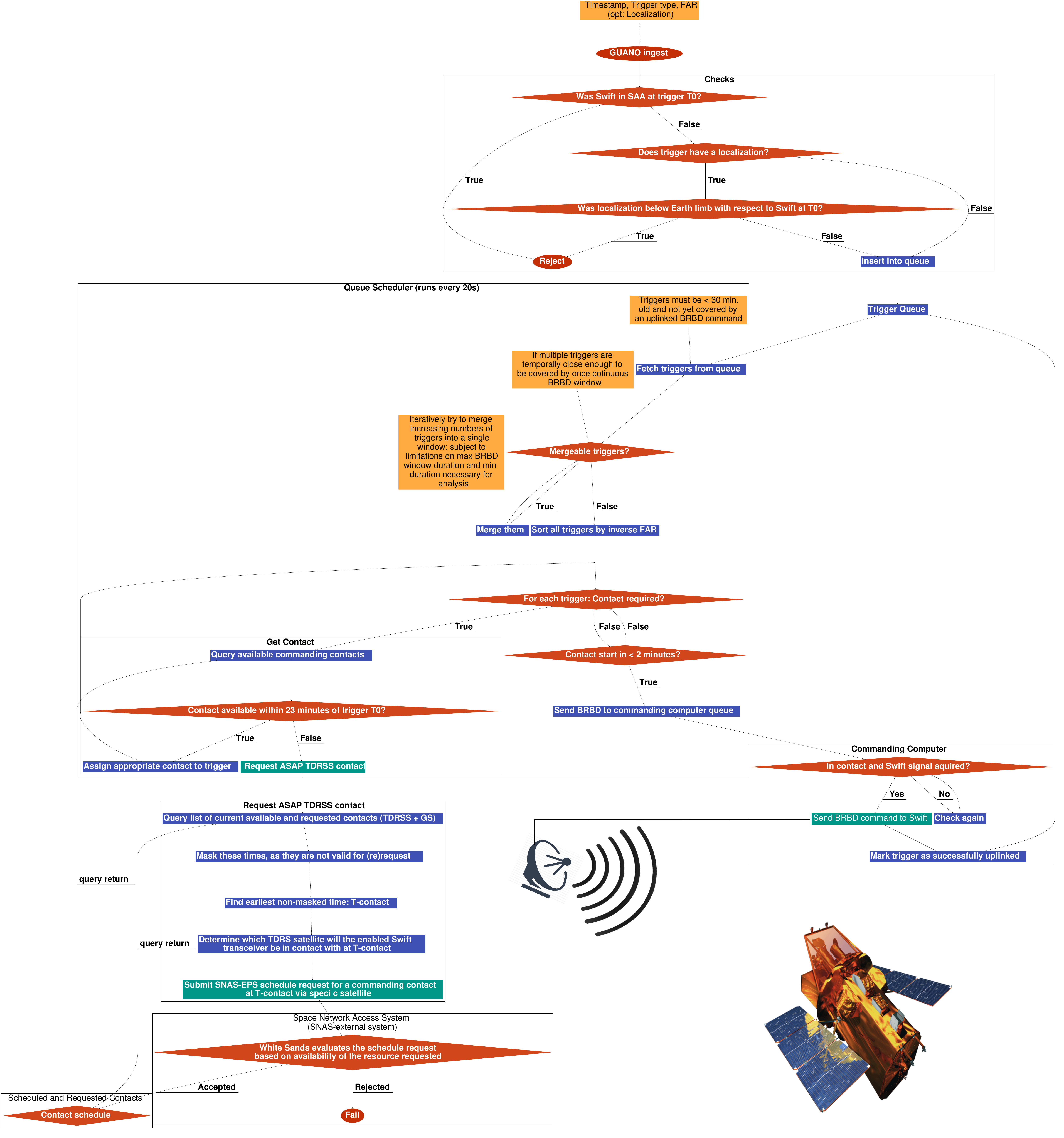}
    \caption{A flowchart depicting the entire GUANO system, from receipt of a trigger to sending a command to \swift\, and its autonomous interactions with external systems such as the Space Network Access System. The entire system runs continuously, and the GUANO queue scheduler fully reevaluates the optimal strategy to ensure data recovery, every 20 seconds, and requests more commanding resources as necessary.}
    \label{fig:flowchart}
\end{figure*}

Here we briefly outline the entire GUANO pipeline, a detailed flowchart depiction of the pipeline and relevant external processes is shown in Figure \ref{fig:flowchart}.

\subsection{The BAT Ring Buffer Dump Command}

The command sent to the spacecraft and BAT to save any existing event data in a certain window around a given timestamp is called a BAT Ring Buffer Dump (BRBD) command. This command has various configurable parameters. For the purposes of the GUANO pipeline, the majority of these parameters are fixed (e.g. events from which parts of the detector array to save, and what types of events) and the only configurable ones are the start time of the requested event window, duration of the requested event window, and which Virtual Channel (VC) in the Solid State Recorder (SSR) the data should be copied to. The Virtual Channel controls the latency/priority of sending the data to the ground.

Given the configurable start time and duration of the requested event window in a BRBD, triggers that are temporally adjacent (but not necessarily related) can be merged by the GUANO pipeline into a single BRBD command, to optimize the use of commanding resources and ensure the recovery of all the relevant data.

The ultimate goal of the GUANO pipeline is to get a BRBD command with the best parameters to the spacecraft, and to execute on the BAT onboard computer, before the relevant data is overwritten in the ring buffer and lost forever. Ensuring the success of this mission, and the safety of \swift\ as a whole in the process, requires a complex real-time system, with several interacting components.

\subsection{Commanding opportunities}
\label{sec:commanding}
As is typical for space telescopes (especially those in LEO), the \swift\ Mission Operations Center (MOC) is not in constant two-way contact with \swift. Sending commands to the spacecraft requires a commanding opportunity, typically called a \textit{contact} or a \textit{pass}. There are two such types of commanding opportunities capable of contacting \swift: Utilizing pre-scheduled passes via one of the ground stations used by  \swift; and on-demand through the Tracking and Data Relay Satellite System (TDRSS). \swift\ performs on average 9 ground station passes per day, which means that less than one in six triggers on average can be successfully commanded using existing ground station passes.

The ability of the \swift\ MOC to autonomously schedule a TDRSS commanding contact on demand, was developed specifically for GUANO, as it is necessary for the recovery of $\sim80$\% of triggers. Previous to this development the latency of manual on-demand TDRSS scheduling and commanding was at minimum 25 minutes, and in reality often much longer, and since it required manual operation, could not be performed with acceptable latency after hours as the \swift\ MOC is only staffed during working hours. This new capability also opens the door to lower latency Target of Opportunity observations with \swift.

However, this comes with a few limitations, firstly the required latency for requesting a TDRSS contact under this system is currently 14 minutes, meaning a contact cannot be requested to begin any sooner than 14 minutes in the future. Secondly, resource demands on the TDRSS network by other users and missions means that occasionally there is no availability of the requested TDRSS resource.

\subsection{Checks}
The extremely low latency commanding required to save the data necessitates an autonomous pipeline. A dedicated listener waits for an event (via either GCN notice or private channel) that meets the triggering criteria. Upon reception it performs a series of checks:
\begin{itemize}
    \item[1.] At the time of the event was \textit{Swift} in the SAA? ($\sim9\%$ of triggers fail this check.)
    \item[2.] If there is a localization associated with the event, was any part of the localization region above the Earth limb with respect to the spacecraft at the time of the trigger (and thus capable of depositing flux onto the detector array). ($\sim30\%$ of triggers with localization information fail this check.)
    \item[3.] If there was a BAT trigger coincident with the trigger time (as BAT will dump event data anyway in this case).
\end{itemize}
If it passes all of these checks, it is approved for commanding, and the trigger is placed into the scheduling queue. 

\subsection{Queue scheduler}
Sending BRBD commands based on astrophysical events requires careful handling of latencies, priorities and overlaps. Each triggering event typically has two parameters that determine its priority. Firstly some measure of the ``goodness" of the event, in most cases this will be a so called "False Alarm Rate" (FAR), typically given in units of Hertz, where a lower FAR is given priority over a higher one. The second prioritization is the trigger time of the event, which currently is only used to determine if an event should be uploaded, based on its likelihood of still having the relevant data in the ring buffer when dumped. 

Trigger handling must be dynamic, and adaptable to changing priorities. For example, when an event occurs, we will associate a pass with it (either a ground station pass, or request a TDRSS contact). However, if a newer, better (e.g. lower FAR) arrives after the first event, but before the scheduled pass, we preferentially should upload that on the pass, and then the first event becomes a secondary priority. A final decision as to what command to what to dump must be made 2 minutes before the start of the commanding pass. As currently only a single BRBD command may be sent per commanding pass, it is important that not only prioritize the best events, but also maximize our chances of dumping all events.

In order to handle events and uploads correctly, we developed a simple queue scheduler for triggers. When there are BRBD events in the queue that have not been uploaded, the queue scheduler first checks whether there is an upcoming pre-scheduled ground-station or TDRSS commanding opportunity that is within the projected ring buffer look back time. If no suitable pass is scheduled, a Tracking Data and Relay Satellite System (TDRSS) forward service is automatically requested via the Space Network Access System (SNAS) EPROM interface to begin in the lowest latency allowable by the Space Network. 

If there is a pass available, either pre-scheduled or requested, on which to upload a BRBD command, the pass is assigned to this BRBD. If the pass is less than 2 minutes away, the queue scheduler creates the most appropriate BRBD command (optimizing the parameters to cover more than one trigger if possible), transfers it to be uploaded to to \swift, and marks the commanding pass as used, and the BRBD entry in the queue as uplinked.

These steps are repeated for all recent triggers ($<30$ minutes old), every 20 seconds. The queue system has no memory of the previous decision that was made, other than requested TDRSS contacts, and the previously uplinked BRBDs, so if more triggers arrive, then their upload strategy is fully re-evaluted to optimize the chance of recovery of all requested data, and if necessary, further commanding contacts will be scheduled to upload multiple commands.

Ring-buffer dumps have a maximum length of 200s, and a default minimum of 90s, so in the case where two events can be covered by a single dump, we merge them together into a single command. Our requirement is that we dump 90s of event data for each GW event (to provide suitable duration of time around a putative short GRB to allow the background to be modelled), so with a maximum dump length of 200s, this means that we can dump two or more events with a single command if the maximum and minimum event times are less than 110s apart.\footnote{This maximum duration of 200s is not necessarily a hard limit and work is ongoing to extend this.} Events that cannot be merged in this way are sent as single BRBD commands.

Scheduling of TDRSS command passes, although automated, can sometimes fail due to issues with limited resources of the Space Network. When a TDRSS pass is scheduled we receive a notification that the scheduling has been successful. However, due to a quirk in the implementation, we do not receive a notification if the scheduling was not successful. However, we have empirically determined that for 90\% of cases, we receive notification of success in $<5$ minutes. If more than 5 mins has passed without notification that the TDRSS pass has been scheduled, the queue scheduler assumes that the pass has failed, and will mark the pass request as timed out. In this case the GUANO scheduler will re-evaluate the upload strategy, and either utilize a ground station pass or request another TDRSS. 

\section{Performance of the GUANO system}
\begin{figure}
    \centering
    \includegraphics[width=0.5\textwidth]{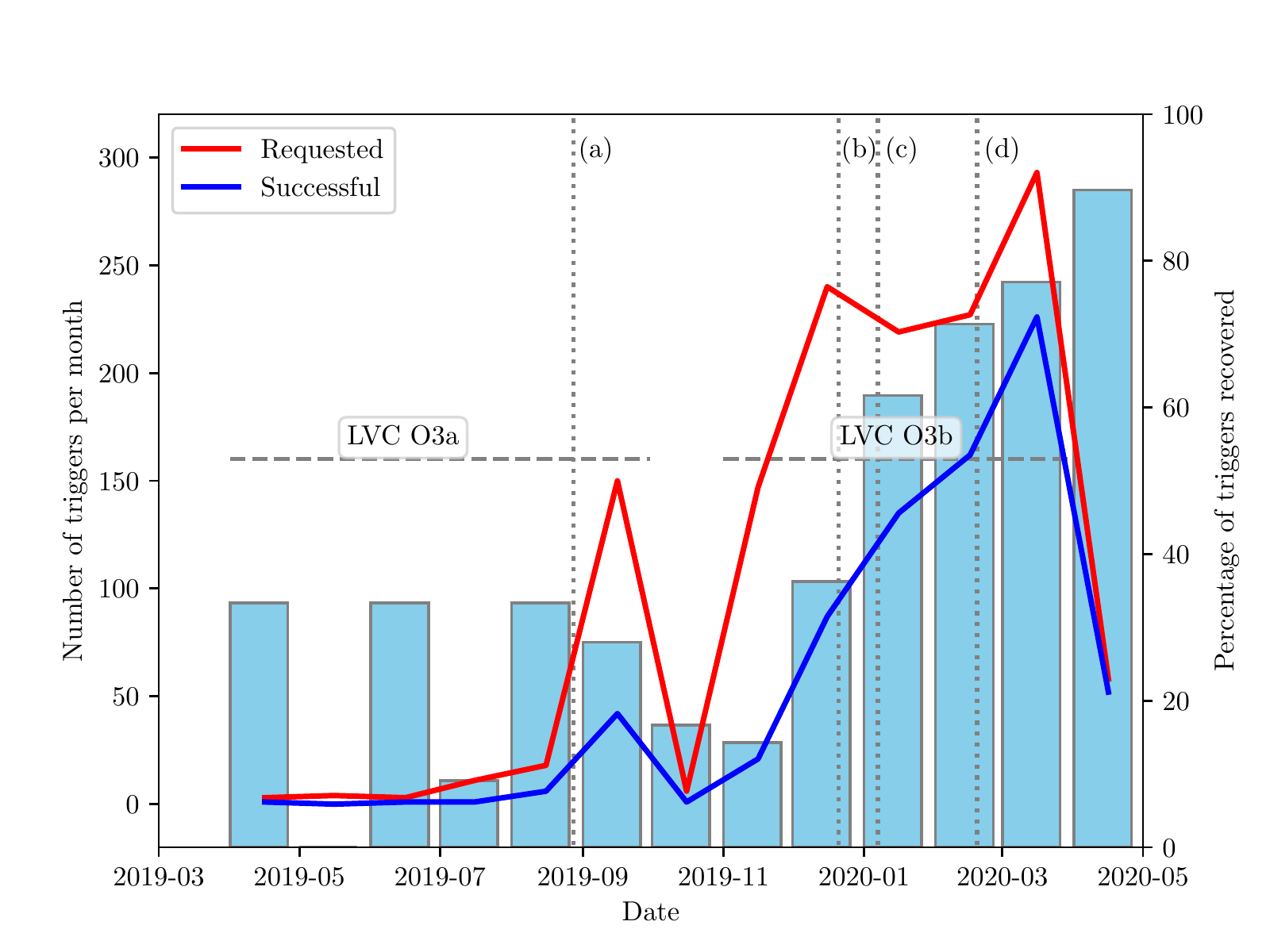}
    \caption{Activity and success rate of GUANO over time. Red and blue lines show the total number of triggers received and the number that were successful respectively per month of operation- and go with the left axis. Blue bars show the monthly percentage recovery rate, i.e. for the number of valid requested triggers, what percentage did we successfully recover BAT event data for? --and go with the right axis. Vertical dotted lines show different epochs of the development timeline: (a) the start of sub-threshold triggers from LVC, (b) when automated TDRSS scheduling came online, (c) when the queue scheduler was brought online and (d) when the TDRSS latency was reduced from 15 mins to 14 mins. Average success rate is now $\sim90\%.$}
    \label{fig:recoveryhist}
\end{figure}

The first successful BRBD command in response to a scientific trigger was sent to \swift\ on April 8, 2019, triggered by the LVC detection of GW candidate S190408an \citep{S190408an}, tentatively classified as a Binary-Black Hole (BBH) merger. At this time the ingestion of triggers was automatic, but the entire scheduling and commanding sequence was manual, resulting in a strong working-hours duty cycle onto the success rate. 
If a trigger arrived during MOC working hours, its associated BRBD had $\sim 70\%$ chance of success, whereas the success rate for after-hours (2/3 of the weekday and 2/7 of the week) was near zero, resulting in a very low average recovery rate. As technical hurdles involved with reducing command latency and automating various parts of the pipeline were overcome, the recovery rate increased apace, eventually reaching near 90\% recovery after the final key components of the GUANO system were implemented in early 2020. An annotated figure showing the recovery rate as a function of time/GUANO development is shown in Figure \ref{fig:recoveryhist}.   

\begin{figure}
    \centering
    \includegraphics[width=0.5\textwidth]{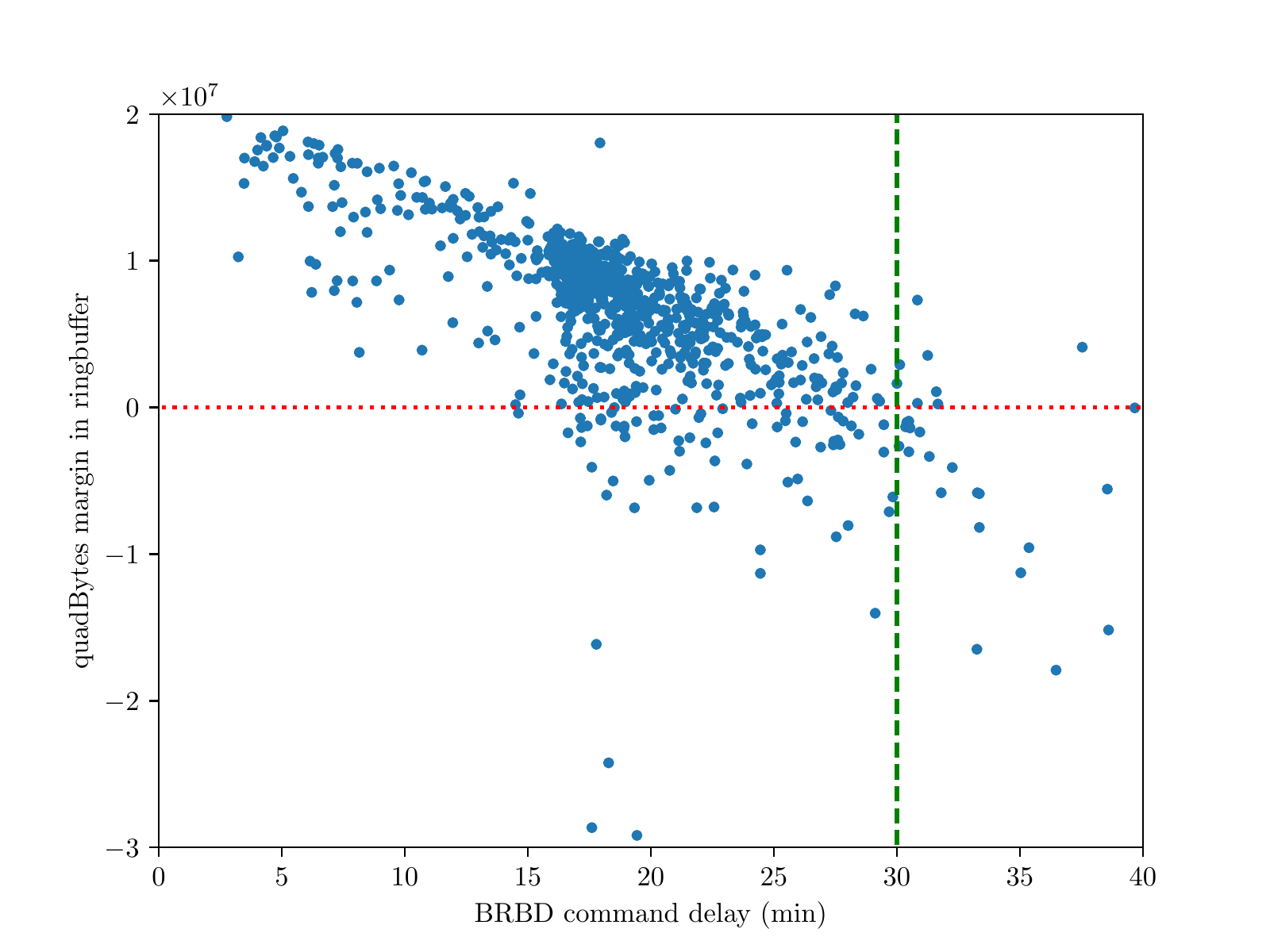}
    \caption{Each blue dot represents a single BRBD command sent to \swift\ by the GUANO system. The horizontal axis denotes how long after the trigger time the command arrived onboard \swift, the total \textit{latency}. The vertical axis shows how much margin remained in the ring buffer before the requested data was overwritten, measured in units of $10^{7}$ quadBytes, at the time the command executed. Negative margins correspond to commands that arrived after the data had already been overwritten, and thus were too late. The red horizontal line at margin=0 marks the Success/Fail boundary. The green dashed line represents the  cut-off time past the trigger time for which we do not attempt an upload. Due to other delays in the system, some BRBDs execute after this cut-off.}
    \label{fig:BRBDlatencymargin}
\end{figure}

While the recovery rate is determined entirely by command latency, the actual latency requirement varies, due to the changing rate of counts hitting the detector and filling up the ring buffer, as described in Section \ref{sec:ringbuffer}. On short timescales, variations in the ringbuffer look back time are strongly dependent on the geographical location of \swift\ in its orbit. Close to the SAA, where the background is higher, the effective duration of the ringbuffer is shorter. In addition, the number of noisy pixels active in the detector also strongly effects the lookback time, and thus infrequent calibration activities on the BAT detector can impact recovery rates for a few hours afterwards. This can be made more clear by examination of Figure \ref{fig:BRBDlatencymargin}. As can be seen, while the majority of commands received onboard with a latency under 20 minutes are uccessful in recovering the data, a small fraction are not, due to the effects discussed above. Commands shown \textit{just} below the red margin line, arrived onboard \swift\ only a few seconds too late to recover the data. The latency cutoff where recovery is successful $99\%$ of the time is $\sim16$ minutes. The cluster of BRBD commands at $\sim17$ minutes, shows the average latency GUANO achieves in its current configuration. We stress that this 17 minutes includes all latencies on the part of both the triggering instrument and the distribution of the notice to the \swift/MOC. GUANO is typically triggered 30-120 seconds after the astrophysical T0, depending on the various latencies of the triggering instrument.  Work is ongoing to continue to decrease the GUANO latency. The lowest achieved latency to date was a BRBD command that was uplinked to \swift\ $\sim2.5$ minutes after the T0 of the astrophysical trigger.

\begin{figure}
    \centering
    \includegraphics[width=0.5\textwidth]{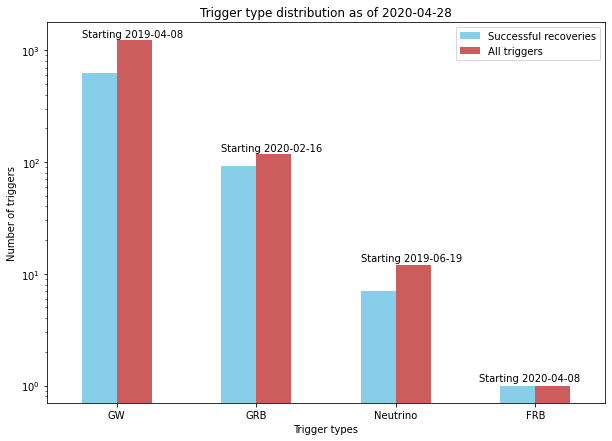}
    \caption{The type distribution and success rate of triggers received by the GUANO system to date. The vast majority of the GW triggers were private sub-threshold triggers from the LIGO/Virgo Collaboration and subject to the \swift/BAT-LVC MoU, which was motivated by the development of GUANO.}
    \label{fig:typedistribution}
\end{figure}

While the GUANO concept was originally motivated by targeted searches for sub-threshold GRBs around the times of GW detections in particular, we eventually opened the system to a larger array of transient types. The type distribution of triggers GUANO has processed and recovered event data for as of publication are shown in Figure \ref{fig:typedistribution}.

We provide a public webpage (\url{https://www.swift.psu.edu/guano/}), that updates live as data are received on the ground, where pointers to the data recovered for each \textit{public} triggering event can be found. 

In the following sections, we provide example results from just a few out of the $>700$ windows of BAT event data around astrophysical triggers recovered by GUANO to date.

\section{LIGO/Virgo unmodelled GW burst candidate S200114f}
\label{section:200114f}

On 2020-01-14 02:08:18.23 UTC, the coherent Wave Burst (cWB) pipeline \citep{cWB} running on real-time data from the Livingston, Hanford, and Virgo detectors of the LIGO/Virgo observatories triggered on an unmodeled transient candidate, S200114f. \citep{S200114f} Notice of this event was received by the \swift\ MOC via private pipeline from the LVC, before public distribution\footnote{Through the low-latency LVC-\swift\ Memorandum-of-Understanding (MoU) pipeline \swift\ receives notice of possible GW events as soon as information is available, to enable GUANO.} of the first notice, and triggered GUANO. The relevant timeline is described in Table \ref{tab:timeline}. Note that the GUANO timeline for S200114f is atypical, as no TDRSS scheduling or commanding was necessary for this event due to the serendipitous temporal coincidence of a ground station with \swift's ground track within the necessary latency window. 

\begin{table}
\label{tab:timeline}
\caption{Timeline of events for S200114f}\vskip -1ex
\begin{tabular}{@{\,}r <{\hskip 2pt} !{\foo} >{\raggedright\arraybackslash}p{5cm}}
\toprule
\addlinespace[1.5ex]
T0 = 02:08:18 & S200114f reaches Earth.\\
T0+$\sim$02:50 & cWB pipeline identifies S200114f in the LVC data stream.\\
%2020-01-14 02:11:13 UTC	moc receives alert
T0 + 02:55 & \swift\ MOC receives alert, and triggers GUANO.\\
T0 + 03:30 & Trigger passes vetting, and is placed in GUANO queue for uplink.\\
T0 + 04:00 & GUANO determines there is a serendipitous ground-station commanding pass within T0+25 minutes, and hence no need for a TDRSS contact. \footnote{If another event had arrived and triggered GUANO while it was waiting to uplink the BRBD command for S20014f, it would then have autonomously scheduled a TDRSS contact, and determined which event to upload first based on their False-Alarm-Rates, see Section 3.}\\
T0 + 24:00 & GUANO passes command to commanding computers for uplink via the Malindi ground station.\\
%2020-01-14 02:34:48
T0 + 26:30 & BRBD command uplinked to spacecraft.\\
%2020-01-14 02:34:49
T0 + 26:31  & BRBD command executes onboard BAT computer, data successfully moved from ring buffer to the Solid State Recorder, and marked for high-priority downlink. \\
T0 + 01:50:00 & Event data arrives on the ground for analysis.\\
\end{tabular}
\end{table}

\begin{figure*}
    \centering
    \includegraphics[width=\textwidth]{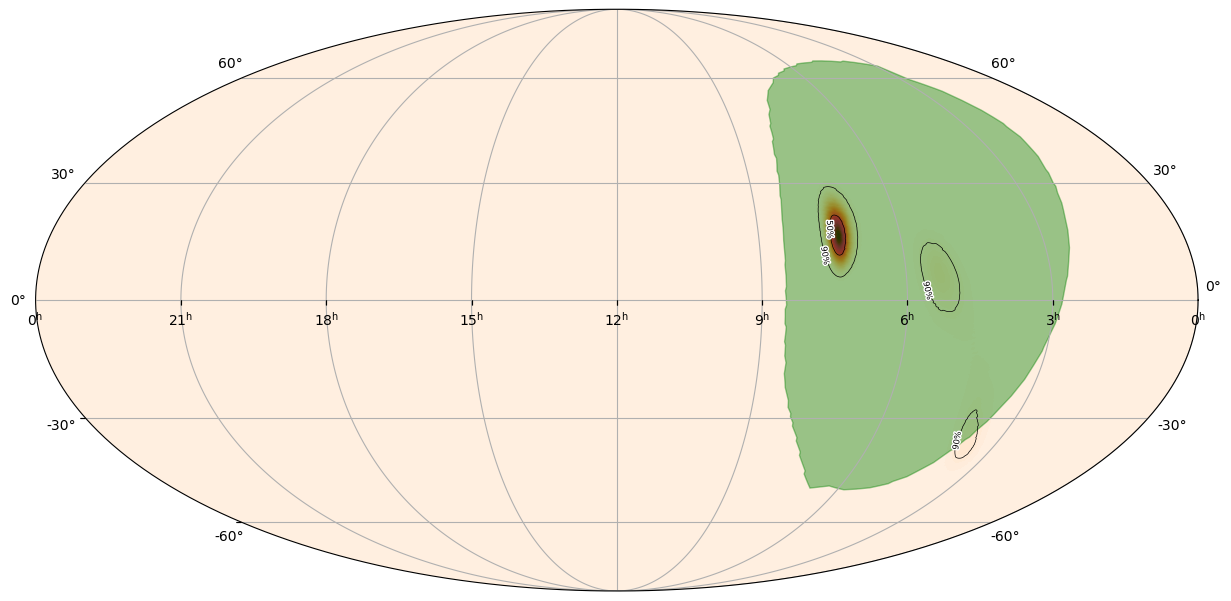}
    \caption{The instantaneous coverage of the \swift/BAT field-of-view at GW T0 is shown in green, along with the localization region, and 50/90 \% containment contours of S200114f in red and black. The BAT field-of-view covered $>99.7\%$ of the GW localization region at T0.}
    \label{fig:coverage}
\end{figure*}

At the time of the GW detection, the \swift\ narrow field instruments (and hence the BAT boresight) were pointed at the \fermi/LAT source 4FGL J0535.3+0934. This source is located within the 90\% localization containment region for S200114f, and thus the BAT field-of-view was covering $>99.7\%$ of the GW localization region at T0, shown in Figure \ref{fig:coverage}.

This high coverage fraction, coupled with the event data saved by the GUANO system, allow very sensitive and complete upper limits to be placed on the existence of prompt GRB-like emission from S200114f. We set such limits below, and throughout also demonstrate the various techniques and results that the event data allow.

\subsection{Detection and localization of weak short duration transients}
\subsubsection{Image analysis results}
With event data available, it's possible to create sky images in specific energy and time ranges. Using the \swift\ FTOOLS\footnote{http://heasarc.gsfc.nasa.gov/ftools} task \texttt{batbinevt}, the event data can be accumulated into detector plane images (DPI), which are used to make sky images with the task \texttt{batfftimage}. When searching for a short transient it can be helpful to create a sky image using a background-subtracted DPI, where the background DPI is made from the event data at times either right before or after the time window of interest. This helps remove coded noise from bright point sources and the diffuse background, but would subtract out signal if the signal was also present during the background time. 
Point sources in the sky images can be searched for using the task \texttt{batcelldetect}, which outputs a catalog of the discovered sources along with several optional outputs including a sky image of the measured Gaussian noise. We use this noise map to set position-dependent fluence upper limits over the duration of the image. 

For S200114f we set upper limits for two durations both in the 14-195 keV energy range, a 10s duration starting 2s prior to the event time and an 88s duration centered on the event time. For the 10s duration we use a background-subtracted DPI to be more sensitive to a transient that starts after the end of the background DPI (10s prior to the event time). For the 88s duration we do not use a background-subtracted DPI so that we do not lose sensitivity to a transient that may have started prior to the earliest data we have available during this pointing.
For each of the durations we find the average upper limit in BAT counts weighted by the localization probability of S200114f at that position in the sky and convert it to a fluence assuming a power law with an index of -1.32 (typical for short GRBs in the BAT band; \citealt{Lien_2016}) using the online WebPIMMS\footnote{https://heasarc.gsfc.nasa.gov/cgi-bin/Tools/w3pimms/w3pimms.pl} tool. We found 8$\sigma$ fluence upper limits over 14-195 keV of $4.1\times10^{-7}$ erg\pcm\ for the 10s duration and $1.1\times10^{-6}$ erg\pcm\ for the 88s duration.

\subsubsection{Max Log Likelihood results}
\label{sec:mll}
The availability of event data on the ground lets us explore analysis methods that take more computational power than can be achieved onboard. One such method is doing a maximum likelihood-based search (for more details see DeLaunay et al. 2020, in prep.). This search uses a Poisson likelihood of the expected counts from background plus a signal model in each of the detectors for several energy bins, and maximizes over the signal model parameters, which include spectral shape, intensity, and sky position. The background model is fit using data outside a temporal search window, here +/-15s around the event time is searched. Inside the temporal search window the search is performed for durations of 0.256s, 0.512s, 1.024s, and 2.048s. The significance of a transient point source is measured using a Test Statistic (TS):
\begin{equation}
    TS = -2\log\bigg[\frac{P(data|H_{S+B})}{P(data|H_B)}\bigg]
\end{equation}
where $H_B$ is the background only model, and $H_{S+B}$ is the best fit signal-plus-background model. The square root of the TS is a comparable measure to a signal-to-noise ratio. 

The sensitivity of this search is a function of the background rates and the source position in the FOV. To find the flux sensitivity to a particular event we inject a simulated signal into the event data and run the search to see if we recover the injected signal above a certain significance. We do this many times at several flux strengths and positions across the FOV. Then, at each position we find at what flux do we recover 90\% of signals at a $\sqrt{TS} > 8$. The signal injections are done at random times inside of the search window and using the detector response for the simulated source position. For the search around S200114f we find the 14-195 keV flux sensitivity assuming a power-law index of -1.32 and averaged over its localization probability for both 0.256s and 1.024s time-scale transients as $2.1\times10^{-7}$ erg\pcm\ s$^{-1}$ and $8.1\times10^{-8}$ erg\pcm\ s$^{-1}$, respectively.

%I'm lazy and don't want to include this section in the paper anymore, it would be a black hole of glitches...

\section{Localizing GRBs discovered by other instruments}
\label{sec:grbselsewhere}
In addition to the utility of the BAT event data to perform the most sensitive searches for GRB emission associated with transients detected in other wavelengths and messengers (as in the previous section), the availability of this data via GUANO also allows BAT to provide arcminute localizations for GRB that otherwise have no, or extremely large, localizations. 

The majority of detected GRBs are reported by \fermi/GBM and INTEGRAL/SPI-ACS. In the case of Fermi-detected GRBs the localizations can range from $\sim100$ sq. degrees to thousands of square degrees. In the case of INTEGRAL/SPI-ACS there is no localization information at all. As a result, without a BAT co-trigger these bursts almost always have no identified afterglows and lack the attendant science that comes with followup observations, including redshift measurement, energetics, jet structure, circum-merger density, etc.

Generally, if a GRB that triggers \fermi/GBM or INTEGRAL/SPI-ACS occurs within the BAT FOV, it also triggers the BAT onboard, with the normal \swift\ GRB response. However, this is not always the case.  While BAT is, a priori, more sensitive then these instruments, and thus regularly triggers on even weaker bursts, a variety of factors (sometimes combined) can result in a GRB detected by other instruments not triggering BAT onboard, even though it originates from within the BAT FOV.
\begin{enumerate}
    \item The BAT onboard algorithms are limited by the necessity to run analyses in real-time on limited processors and thus are not able to reach 100\% of the recovery potential that can be achieved on the ground with the same techniques.
    \item The BAT onboard algorithms do not run while \swift\ is slewing, which is $\sim15$\% of the time.
    \item The GRB, while it may be intrinsically bright, is incident upon BAT from a line-of-sight for which the BAT coding fraction is very low (read: near the edge of the field-of-view), and undetectable using the conventional BAT analysis techniques, necessitating instead complex analyses such as the max-log likelihood search described in Section \ref{sec:mll}.
    \item The GRB may be extremely spectrally hard, with very weak emission in the BAT bandpass.
\end{enumerate}
By recovering BAT event data around the times of GRBs discovered by other missions, we can close this gap when possible and fully exploit the localizing power of BAT.

Since opening the GUANO listener to GRBs in mid-February 2020, GUANO has recovered BAT event data around the time of a large majority of the GRB triggers from \fermi/GBM, INTEGRAL\footnote{GUANO also triggers on the weak/sub-threshold stream from INTEGRAL/ISGRI, whose astrophysical purity is unknown \citep{INTEGRAL-weak}. None of these triggers have been seen in BAT to date.}, and CALET that do not also trigger BAT. Of these, arcminute localizations for 4 GRBs have been recovered using the BAT event data provided by GUANO, at a rate of approximately 2 per month: GRBs 200216A \citep{GRB200216A-GUANO}, 200228A \citep{GRB200228A-GUANO}, 200325A \citep{GRB200325a-GUANO}, 200405B \citep{GRB200405B-GUANO}. This rate, and the analysis of historical BAT FOV overlaps with GBM localizations, leads to the conclusion that GUANO can effectively increase the worldwide net rate of arcminute localized GRBs by at least $15$ GRBs per year, an increase of $17\%$ over the current rate. 
Of these 4 GRBs, the reasons they did not trigger BAT onboard were the following:
\begin{itemize}
    \item One because \swift\ was slewing at the time of the GRB.
    \item One due to an inefficiency in the onboard trigger algorithms. In this case, the BAT onboard algorithms only made, and searched for sources in, images in the 100-350 keV bandpass around the time of this burst, whereas the source was found on the ground when including lower energy events.
    \item Two because they originated from a location on the sky with a very low partial coding fraction, and thus were too intrinsically weak in the sky images to be found using the standard BAT analysis, both onboard and on the ground.
\end{itemize} 

Here we outline the GUANO enabled localization of one of these bursts, short hard GRB 200325A:

\subsection{GRB 200325A}
GRB 200325A was discovered in real-time by \fermi/GBM \citep{GRB200325a-Fermi-real}, INTEGRAL/SPI-ACS, and AGILE/MCAL \citep{GRB200325a-AGILE} with T0 of 2020-03-25 03:18:31.7 UT. Detections were also later reported by AstroSat CZTI \citep{GRB200325a-AstroSat}, Konus-WIND, GRS-Odyssey, and HEND-Odyssey \citep{GRB200325a-IPN}. The only of these instruments capable of independently localizing the GRB was \fermi/GBM, which provided a 3-sigma containment region of $\sim800$ square degrees. \swift/BAT did not trigger on this burst.

The low-latency notices of this GRB from \fermi\ (distributed at T0+8s) and INTEGRAL (distributed at T0+44s) both triggered GUANO. The GUANO queue scheduler merged these events into a single specific BRBD command, requesting 200s of event data from [-50,+150] around T0, and ensured its uplink by scheduling and confirming an on-demand TDRSS contact within the required latency window. GUANO then sent the BRBD for uplink to \swift. The BRBD command executed onboard the BAT computer at T0+13 minutes, moving the requested data from the ring buffer to the solid-state recorder, and marking it for high priority downlink. The requested data arrived on the ground at T0+46 minutes.

\begin{figure}
    \centering
    \includegraphics[width=0.5\textwidth]{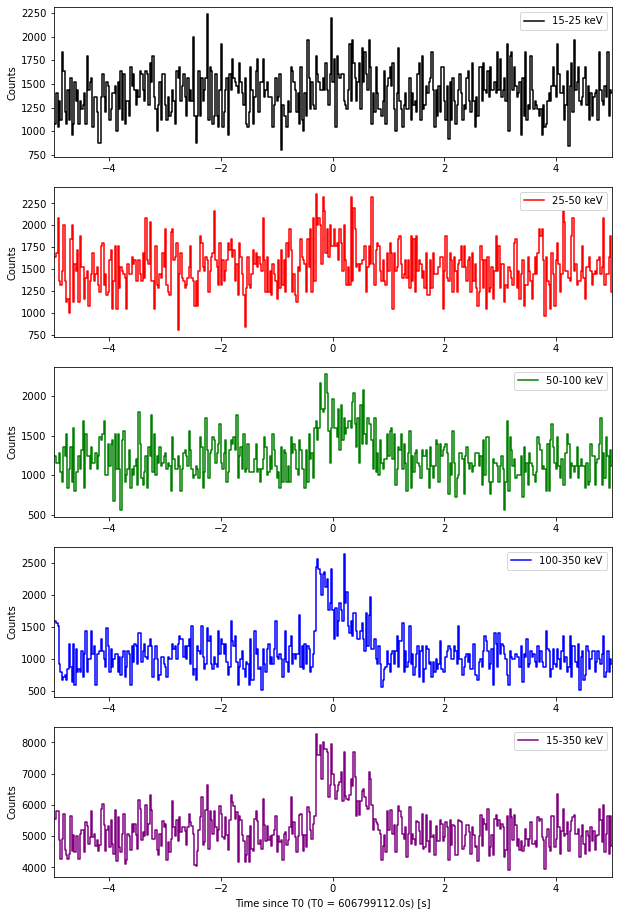}
    \caption{A 25-ms binned full detector summed lightcurve of GRB200325A, constructed from the event data recovered by GUANO. Because the GRB originated at such low partial coding, the full detector summed light curve shown here (not background subtracted) is much more significant than the mask-weighted light curve (background subtracted), as more counts from this GRB arrived at the detector after penetrating the instrument shield through the sides, than those that passed through the mask openings.}
    \label{fig:GRB200325A-lc}
\end{figure}

We used the \swift\ FTOOLS task \texttt{batbinevt}, to accumulate the event data into DPIs, and then made these into sky images with the task \texttt{batfftimage}. The sky images were created using a background-subtracted DPI using event data from directly before the GRB interval. The sky images were then processed with the task \texttt{batcelldetect}, to search for sources. The DETECTION mask was used, in order to search out to the largest solid angles/lowest possible partial coding. Successive manual trials were performed, optimizing both the `source' and `background' intervals, as well as the energy range, for the DPI and sky image accumulation, until we were able to produce a sky image with a source $>7$ sigma. The burst location was found in an image made with an interval from -0.064s to 1.024s, and with events in the energy range 20.0 - 195 keV, with an SNR of 7.5 from \texttt{batcelldetect} and at an extremely low partial coding fraction of 8.2\%. This very low partial coding, and thus the need to fine tune the source interval and energy ranges for a construction of a sky image with an acceptable SNR, explain fully why the BAT onboard algorithms did not trigger on this burst. However, the repeated optimization of the images induced significant trials onto the detection.

\begin{figure}
    \centering
    \includegraphics[width=0.5\textwidth]{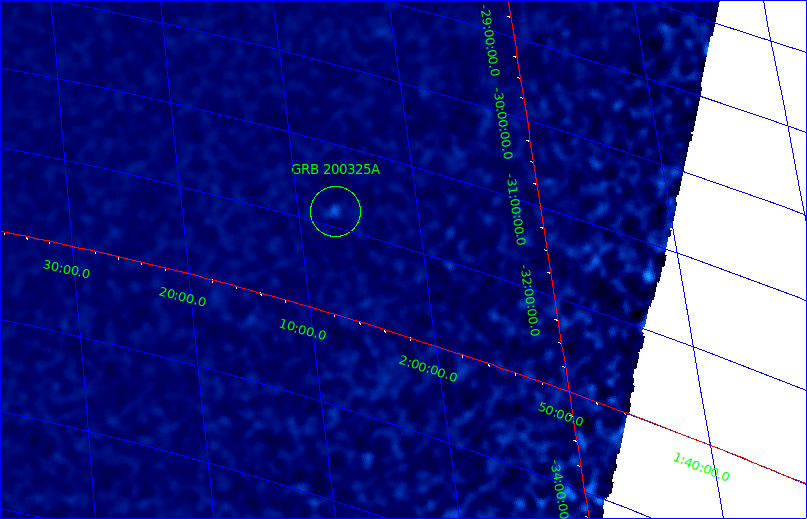}
    \caption{A BAT hard X-ray sky image of GRB 200325A, showing its proximity to the absolute edge of the field of view and the growth of image noise close to the edge. For such sources, and weak sources generally, the max-log likelihood analysis described in Section \ref{sec:mll} is required to distinguish real astrophysical sources from noise.}
    \label{fig:my_label}
\end{figure}

Simultaneous to the image analysis, the computationally expensive maximum-likelihood based search (DeLaunay et al. 2020, in prep.) was also run on the BAT event data at the time of this burst. This search returned the same source as the image analysis, with no iteration or fine-tuning required, and at a much higher significance with a $\sqrt{TS}$ of 21.7. 

We distributed a GCN circular reporting the localization to the community \citep{GRB200325a-GUANO}. The burst coordinates are 31.7203, -31.816 with a 90\% containment on the uncertainty (systematic plus statistical) of 4 arcminutes. Unfortunately, this location was too close to the Sun to allow any followup to search for an afterglow.

\begin{figure*}
    \centering
    \includegraphics[width=\textwidth]{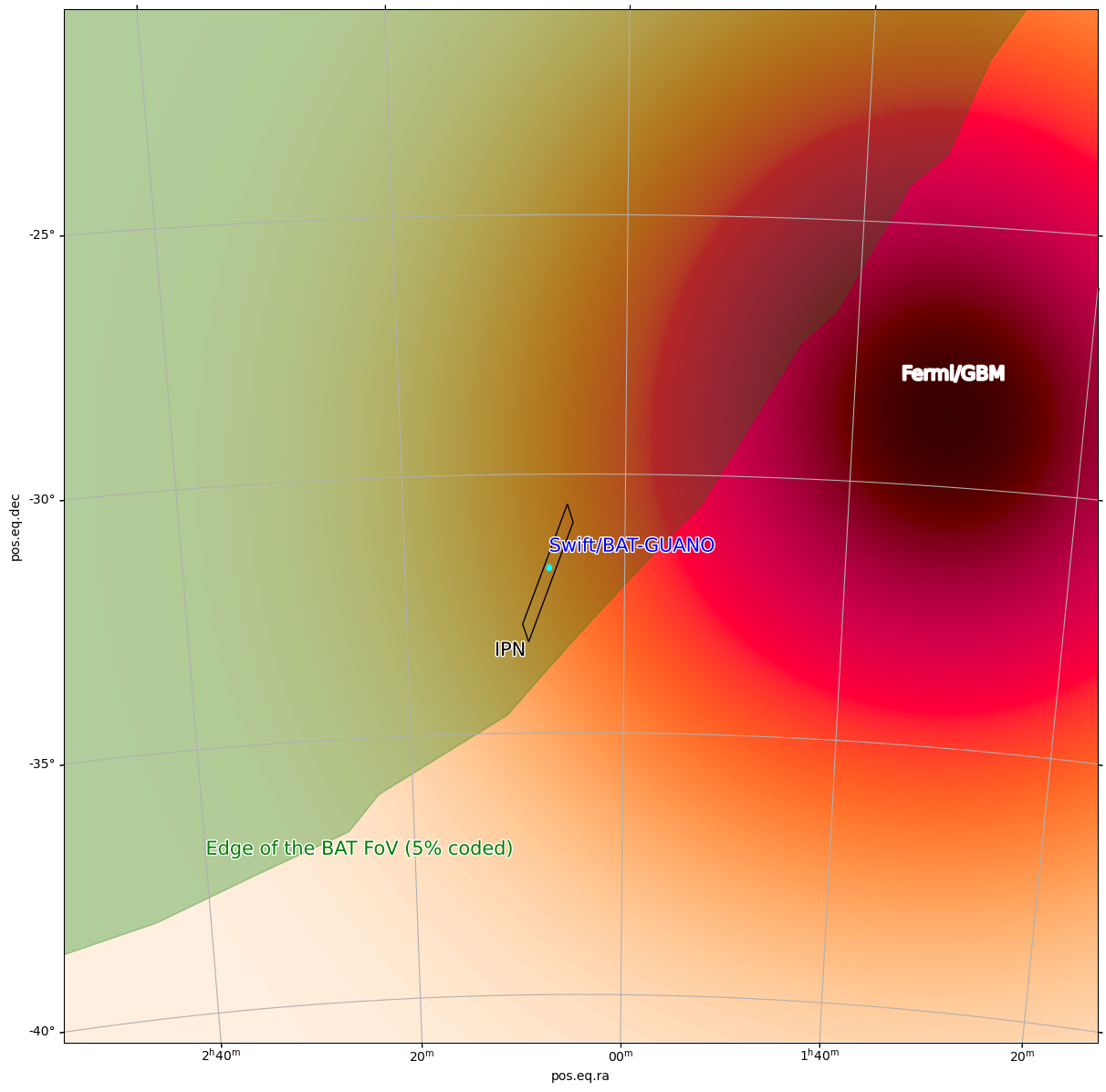}
    \caption{Skyplot of the localizations for GRB 200325A. The black-red-orange heatmap shows the \fermi/GBM localization region, the black polygon shows the IPN timing localization, the small cyan circle shows the \swift/BAT-GUANO localization we reported $\sim$1 day before the IPN localization was announced. The green shaded regions shows the extent of the BAT field-of-view (down to 5\% coding) at T0.}
    \label{fig:GRB200325A}
\end{figure*}

Approximately a day after we reported this arcminute localization, the Inter-Planetary Network, localized the burst via intersecting timing annuli with KONUS-Wind, Mars Odyssey-HEND, \fermi/GBM, and INTEGRAL SPI-ACS. Their 90\% localization region spanned 2023 square arcminutes \citep{GRB200325a-IPN}, and agreed with the \swift/BAT-GUANO derived position. The various localizations for this burst are shown in Figure \ref{fig:GRB200325A}.

\subsection{Non-imaging localizations and out-of-FOV science}
The event data recovered by GUANO can also be used to help localize a GRB even if that GRB does not originate from within the BAT coded field-of-view. High-energy photons from GRBs originating from anywhere on the unocculted sky (in Low Earth Orbit $\sim1/3$ of the sky is occulted by the Earth) can penetrate through the Z-shield surrounding the BAT instrument, or even through the entire spacecraft body, and deposit counts into the BAT detector.  Indeed, $\sim40\%$ of IPN-reported (read: bright) GRBs are also found in searches of BAT rates light curves, but originate from outside the coded field-of-view. With some notable exceptions (eg the magnetar hyperflare of SGR 1806-20, one of the earliest bursts of the \swift\ mission; \citealt{Palmer05}) these rates data for out of FOV GRBs have typically not been of particular scientific use, aside from confirming the reality of a detection, for the following reasons:
\begin{enumerate}
    \item The rates data (max temporal resolution 64 ms) is not normally of high enough temporal resolution to allow BAT to participate in IPN timing localizations.
    \item The rates data is only available in 4, broad, energy bands.
    \item The lack of calibrated responses for out-of-FOV lines of sight typically preclude using these data for spectral analyses or any other type of analysis beyond crude examinations of the light-curve morphology.
\end{enumerate}

The on-demand event data from GUANO effectively solves the first two of these issues. Indeed, the GUANO-enabled event data has already allowed BAT to participate in IPN timing localizations of several GRBs (eg \citealt{GRB200405B-IPN, GRB200415A-IPN}). In addition, the ability to clean noisy/hot detectors and glitches out of the rates data using the event data from GUANO allows for more confident claims of detection of out-of-FOV bursts, and can thus allow BAT to reduce the localization region of \fermi/GBM bursts by ruling out parts of the sky that were occulted for \swift\, as in the case of GRB 200307A \citep{GRB200307A}.

The converse is also true, a non-detection of a GRB in the ground analysis of GUANO-derived event data can rule out the GRB's origin from within the BAT coded FOV. The BAT FOV often overlaps substantial fractions of the \fermi/GBM localization, even for bursts that BAT does not trigger on. For some of these (ie GRB 200325A, above, and others) the burst is eventually found within the BAT FOV. However, in cases where the source is not found using the powerful ground analyses, this overlap region can be ruled out and the size of the localization for the burst effectively reduced, often dramatically. From simulations and an analysis of archival data, we predict using such a technique BAT can reduce the size of \fermi/GBM localizations by 50\% for at least $\sim15$ GRBs per year. 

With regard to Item 3: The event data from GUANO could even be used to characterize, validate, and build response functions for BAT along many different out of FOV lines-of-sight, and thus allow this data to be used in the future for spectral fits, or even a rough localization in the scheme of eg \fermi/GBM Team \citep{goldstein2020} (where an astrophysical spectrum is assumed, and then the localization fit from that) or BALROG \citep{balrog} (where the spectrum and localization are fit simultaneously). We note that even for lines-of-sight $>100$ degrees away from the BAT boresight ($>50$ degrees outside the coded FOV), the BAT effective area is still hundreds of $\text{cm}^2$ at energies $>100$ keV \citep{Palmer05}, comparable to the effective area of \fermi/GBM. Extending the calibrated responses out to such angles would result in BAT data products (eg spectra, possible rough localizations) being usable, beyond just the timing localization and light curve morphology described above, for many more GRBs per year.

A possible approach would be to take the known spectrum of a \fermi/GBM burst, that is seen in BAT from outside the FOV, and jointly fit that spectrum with the observed spectrum from the GUANO event data, and thus generate a response function for that particular line-of-sight, effectively building a response in a data-driven way, and iterate for each line-of-sight for which a GRB exists. Alternatively the data could be used to validate the predicted response from the Swift Mass Model, a pre-launch GEANT4 model of the BAT instrument and the \swift\ spacecraft body that was used in Monte Carlo simulations to generate Detector Response Matrices for BAT pre-launch, but whose out-of-FOV responses have not been extensively validated due to the previous absence of data.
% Likelihood of SwiMM being retrievable at this point in time - ZERO :) -Jamie

Such a project is large and well outside of the scope of this work, but we comment that GUANO dumped event data for GRBs originating from outside of the FOV, but with known spectra from other instruments (eg \fermi/GBM) are now accumulating at the rate of approximately a few lines-of-sight per week, already having accumulated many such bursts as of publication. The \fermi/GBM response is sampled from 272 lines-of-sight \citep{Connaughton15}.

\section{Conclusions}
The ability to recover event data from the BAT instrument on demand significantly increases its sensitivity to weak transients that do not trigger onboard, as would be the case for an off-axis GRB at typical BNS ranges achievable by the ground-based gravitational wave interferometers, effectively increasing the rate of detections and arcminute localizations of GRB 170817-like bursts by $>400\%$. The data can be exploited in various ways to accomplish this, as they enable the creation and search of background subtracted gamma-ray sky images, the use of new statistical techniques designed to fully exploit the latent information associated with each individual count, and the confident identification, classification, and removal of glitches and GRB-mimickers from the data. 

Using just a small fraction ($\sim200$ s) of the total BAT event data recovered by the GUANO pipeline ($\sim75$ Ks) to date, we demonstrate its utility directly by providing the deepest upper limits on prompt GRB-like emission associated with the GW burst candidate S200114f, and show how the data can be used to recover an arcminute localization for GRBs triggered by other missions, such as short GRB 200325A. We provide a public website\footnote{\url{https://www.swift.psu.edu/guano}} that records and reports the event data saved in response to \textit{public} triggers, and makes this data fully available to the community for use. Indeed the data has already seen use by the broader community, beyond what it was designed for, in the Inter-Planetary Network.

In addition, the novel operational capabilities described here, developed as they were necessary to recover this data, demonstrate the first fully autonomous, on-demand, extremely low latency commanding of a space telescope based on astrophysical triggers. These capabilities open the door to other high-impact science, including fully autonomous extremely low latency repointing of \swift\ for Target-of-Opportunity observations with its narrow-field instruments. 

% I wrote the next few paragraphs - Jamie
The development of the capabilities demonstrated by GUANO could have a profound effect on the operational capabilities of future space missions, as well as enhancing the science capabilities of \swift. For example, when developing new missions, the cost and associated risk of creating novel flight software to make spacecraft react autonomously to transient triggers is often outside the scope and budget allowed. In addition the costs of running Mission Operations for future missions is significantly higher if out-of-hours human response is required. By developing a system to both automate the ingestion, rank ordering and validating of transient triggers from multiple sources, combined with an automated way to generate and send the relevant commands to respond to those events to the spacecraft, it now becomes possible to build 24/7 response to transients into low-cost missions, as well as larger missions concepts with ultra-low latency ToO requirements.  

We encourage TDRSS to develop a truly on-demand commanding capability, similar to what exists from commercial providers (eg GlobalStar or Iridium), the commanding equivalent of the TDRSS Demand Access System (DAS) that already exists and is regularly utilized by \swift\ for return service. Such a capacity is also necessary for strong science cases demanding commanding with latency of order seconds (Tohuvavohu et al. 2020c, in prep.), as compared to the $\sim$ minutes latency described here. Absent such a capability from TDRSS, we remark that it may behoove designers of next-gen extremely rapid response space telescopes to consider the use of a commercial network for their ToO, or other extreme-low-latency, commanding. 

The development of GUANO for \swift\ serves as a flight-proven retirement of the risk associated with fully autonomous space telescope commanding, and thus opens the door for such capabilities to be included in the design of future missions. In the upcoming era of high-transient detection rates from the likes of the Vera C. Rubin Observatory \citep{LSST}, fully autonomous and rapid transient response and follow-up will become more crucial than ever. 

\section*{Acknowledgements}
We thank the \swift\ Flight Operations engineers for their invaluable assistance bringing GUANO online, and for consistent safe stewardship of the \swift\ mission. We acknowledge and thank Eric Siskind for providing critical guidance on autonomous TDRSS scheduling. We thank the \swift\ Observatory Duty Scientists for their support while this program was in its manual testing phases. We acknowledge Dustin Lang and Amanda Cook for their help naming the project. We thank Eric Burns for helpful discussions regarding \fermi/GBM data.

\vspace{2\baselineskip}
\textit{Facilities}: Neil Gehrels \swift\ Observatory
\vspace{\baselineskip}

\textit{Software}: healpy \citep{healpy}, Matplotlib \citep{matplotlib}, ligo.skymap \citep{ligo.skymap}, NumPy \citep{numpy}

\bibliographystyle{aasjournal}
\bibliography{main}

\end{document}